\documentclass[11pt, letterpaper]{article}
\usepackage[utf8]{inputenc}
\usepackage[margin=1.5cm]{geometry}
\usepackage{titlesec}
\usepackage{tabu}
\usepackage{enumitem}
\usepackage{amssymb}
\usepackage{xcolor}
\newlist{selectlist}{itemize}{2}
\setlist[selectlist]{label=$\square$,leftmargin=*,noitemsep,topsep=0pt}

\usepackage{graphicx}
\usepackage{subfig}
\usepackage{caption}
\usepackage{cite}
\usepackage{mwe}
\usepackage{subfig}
\usepackage{blindtext}
\usepackage{lmodern}

\usepackage{hyperref}
\hypersetup{
    colorlinks=true,
    linkcolor=blue,
    filecolor=magenta,      
    urlcolor=blue,
}
 
\titleformat{\section}[block]{\hspace{1em}\bfseries}{\thesection.}{0.5em}{} 
\titleformat{\subsection}[block]{\hspace{1em}}{\thesubsection}{0.5em}{}

\begin{document}

\begin{flushleft}

\setlength{\parindent}{0pt}
\setlength{\parskip}{10pt}

\textbf{Article title}\\ Dark-field light scattering microscope with focus stabilization

\textbf{Authors}\\ Anna Peters*, Zhu Zhang, Sanli Faez

\textbf{Affiliations}\\ Nanophotonics, Debye Institute for Nanomaterials Science, Utrecht University, 3584CC Utrecht, The Netherlands

\textbf{Corresponding author’s email address and Twitter handle}\\ a.peters1@uu.nl

\textbf{Abstract}\\ We present detailed design and operation instructions for a single-objective inverted microscope. Our design is suitable for two dark-field modes of operation: 1- total internal reflection scattering, and 2- cross-polarization backscattering. The user can switch between the two modes by exchanging one mode-steering element, which is also adapted to the Thorlabs cage system. To establish a stable background speckle for differential microscopy the imaging plane is stabilized with active feedback. We validate the stabilization efficacy by performing long-term scattering measurement on single nanoparticles. This setup can be extended for simultaneous scattering, fluorescence, and confocal imaging modes.


\newpage
\textbf{Specifications table}\\
\vskip 0.2cm
\tabulinesep=1ex
\begin{tabu} to \linewidth {|X|X[3,l]|}
\hline  \textbf{Hardware name} & Single-Objective stabLe Inverted Dark-field microscope (SOLID)
  \\
  \hline \textbf{Subject area} & 
  \vskip 0.1cm
  \begin{itemize}[noitemsep, topsep=0pt]
  \item Coherent imaging
  \item Engineering and Material Science
  \item Analytical Chemistry
  \end{itemize}
  \\
  \hline \textbf{Hardware type} &
  \vskip 0.1cm  
  \begin{itemize}[noitemsep, topsep=0pt]
  \item Imaging tools
  \end{itemize}
  \\ 
\hline \textbf{Closest commercial analog} &
  While some commercial microscopes can be customized to provide this imaging modality, we are not aware of a system that provides cross-polarization mode in reflection.
  \\
\hline \textbf{Open source license} &
  CC BY-SA 4.0
  \\
\hline \textbf{Cost of hardware} &
  38,500 EUR up to 59,541 EUR (with Hamamatsu Orca Flash 4.0 sCMOS camera)
  \\
\hline \textbf{Source file repository} & 
  \url{https://doi.org/10.17605/osf.io/t3yvd}
  \\
\hline \textbf{OSHWA certification UID} \vskip 0.1cm &
NL000012
\\\hline
\end{tabu}
\end{flushleft}

\newpage
\section{Hardware in context}
Optical detection of particles smaller than one micrometers has become relevant in many analytical methods in medicine, nanomaterials fabrication, and the cosmetic industry. 
While electron microscopy is widely used for specialized investigations of particles below one micrometer, optical methods are preferred for desktop applications and point-of-care investigations. 
In dark-field microscopy, unlabeled nanoparticles can be visualized under an optical microscope\cite{meng2021micromirror} even inside a liquid. By monitoring the nanoparticle dynamics one can estimate and further manipulate their characteristics, e.g. the charge state, binding with the environment or adhesion on a substrate\cite{wang2019before}, and their composition.
Small particles are difficult to detect with coherent scattering as the scattered intensity scales with the sixth power of the particles' radius. Regarding biological samples, it becomes more challenging due to their low refractive index, heterogeneous distribution in the optical field, and low scattering cross-section. 
However, interferometric-enhanced methods such as iSCAT \cite{ortega2016interferometric,taylor2019interferometric,liebel2017ultrasensitive,jollans2019nonfluorescent} and phase-contrast imaging \cite{matlock2017differential} offer high sensitivity. 
In some coherent methods, the properties of the reference beam can be altered, which leads to a higher signal-to-noise ratio (SNR) \cite{daaboul2010high}. 
\\\\
Dark-field microscopy gains popularity as a background-free detection tool, where small, bright features appear on a dark background. 
However, methods based on scattering are inherently more sensitive to speckle creation by partial reflection from each optical elements. 
This background speckle is an interference pattern that can rapidly vary due to drift in the optical path and the illumination laser frequency.
For this reason, scattering-based optical detection tools often require a more stable and robust set-up than methods based on fluorescence. 
Therefore, mechanical vibrations and other drifts, such as fluctuations in the laser mode or convection in optical light path, must be minimized or eliminated. 
\\\\
While several groups \cite{kwakwa2016easystorm,alsamsam2022mieye, merces2021incubot,maia2017100,meloni20173d,guver2019low} have reported their open-source designs for fluorescent-based microscopy methods, to our knowledge, no coherent imaging microscope with open hardware have been implemented with the specification required for stable imaging of highly sub-wavelength nanoparticles.

\section{Hardware description}\label{hardware_description}

Here we present an inverted reflection-based dark-field microscope with two modes of operation.
These two modes are 1- total internal reflection scattering and 2- cross-polarization imaging, which can be applied inside the same skeleton by replacing a single part: a compound beam-directing element. 
To ensure high sensitivity, we need a robust skeleton with few moving parts and solid-on-solid contacts in between the individual components. 
The fundamental design concept is inspired by the Thorlabs cage system; a cost-effective and accessible solution for coherent optical setups. 
Our customised set-up can be extended for fluorescence imaging, but we will not further discuss this application in this paper. 
\\\\
The setup, depicted in Fig \ref{top_view}, is an inverted dark-field microscope in reflection mode, where the sample is illuminated through a thin cover glass. 
The sample of choice can be a solid, e.g. small features on microscope coverslips or thin films, or particle suspensions sandwiched between two glass slides. 
The sample holder is made from aluminium (see \nameref{sample_holder}) and adapted to the exact size of the sample (usually 24x40 $mm$ coverslips). 
The sample is additionally fixed with magnets on top to minimize the drift when positioned on the oil-immersion objective, while making sample exchange easy and fast. 
The fiber-coupled illumination beam is directed onto the dichroic mirror (Component No.12), where the short wavelength of 488 $nm$ is reflected. The blue laser hits the sample from below, perpendicular to the object's surface plane.
The scattered light is collected through the high NA objective and the imaging optics onto a scientific CMOS camera. 
While operating in cross-polarization imaging, a second laser beam at 780 $nm$ (Component No.0) needs to be installed for optical feedback control. The reflected infrared laser beam is guided onto the QPD (Component No.56). 
The difference signal on the QPD is recorded is used as feedback for correcting the drift of the glass slide by the piezo-actuated stage (Component No.27). 
In the following section we walk through the setup parts and discuss them with more details.
\subsubsection*{Total internal reflection-based dark field microscope}
In total internal reflection illumination, the incident beam hits the glass-sample interface under an oblique angle larger than the critical angle (Glass-Air interface: 42 degrees and Glass-Water interface:  62.45 degrees). 
The back-reflected light leaves the interface under the same angle.
Most of the scattered light propagates towards the higher-index side of the interface. 
Here, the specularly-reflected beam is used as the optical feedback as its direction is highly sensitive to displacement of the glass interface.
The central beam-stirring in this setup is a custom-designed element, matching the cage system, that keeps the micromirrors behind the objective;  see Fig \ref{micro_arm}. 
Each of the arms is mounted on a kinematic stage, which provides two degrees of freedom necessary for precisely aligning the illumination and reflection paths. 
To connect this customized part to the Thorlabs cage system, we used the base mount (Component No.35), providing another (rotational) degree of freedom for both arms. 
During alignment, we need to make sure that the maximal intensity is hitting the first gold mirror to have the greatest laser-particle interaction, hence the largest scattering signal on the camera, and afterwards ensure that the maximal intensity is hitting the second gold mirror to read the highest possible value on the QPD for feedback stabilization, and to remove the specular reflection from the imaging path, as much as possible.

\subsubsection*{Cross polarization-based dark field microscope}

\begin{figure}[htp]
    \centering
    \includegraphics[scale=0.30]{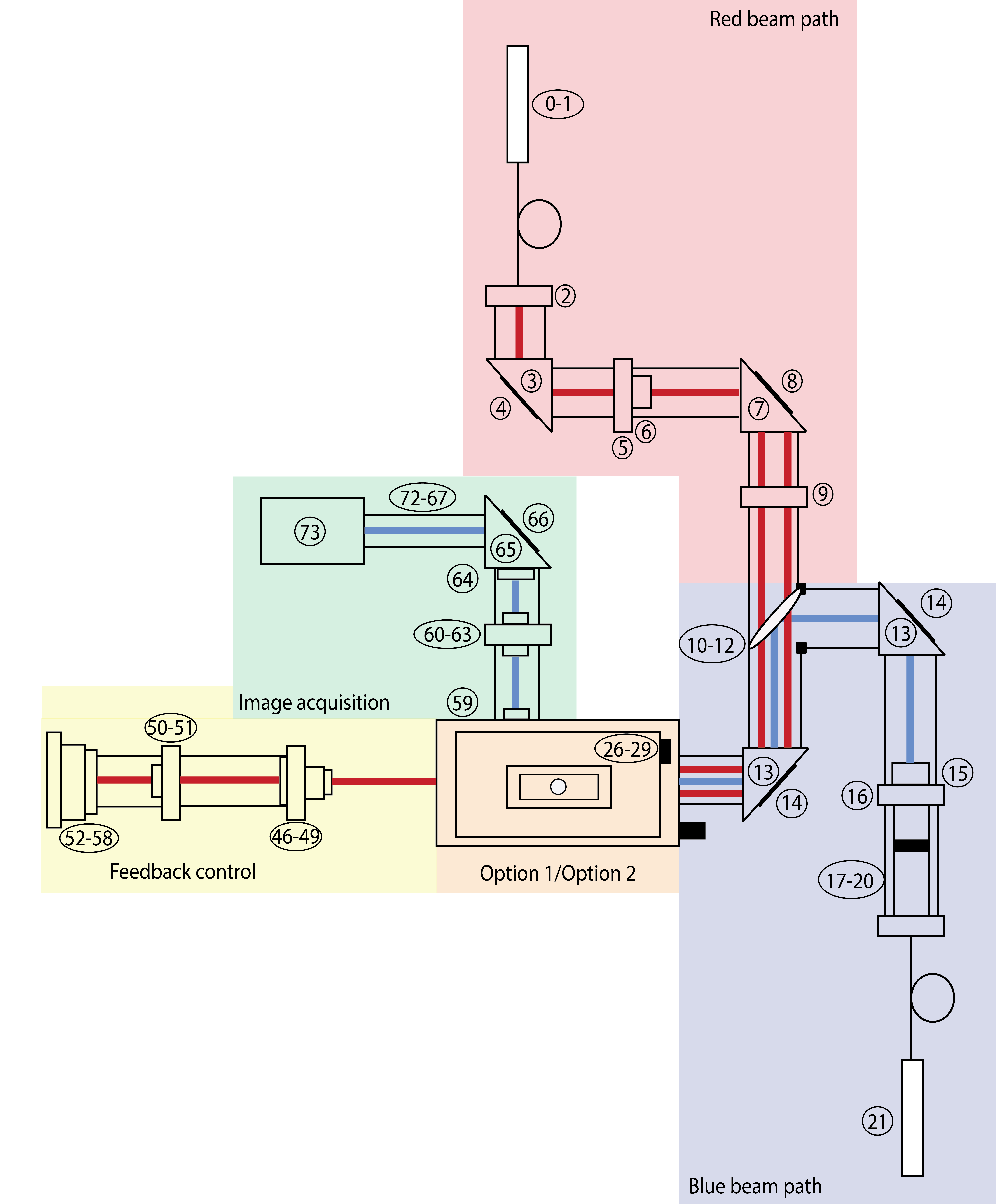}
    \caption{Top view of experimental setup. The illumination beam path is shown in the blue colored region. The beam hits the sample from below (orange colored region) and the reflected beam is guided further to the camera (green colored region). If you prefer to work with the total internal reflection imaging setting, you do not need to install the red beam path. The reflected blue beam acts as the optical feedback and hits the quadrant photodiode (yellow colored region). While working with the cross-polarization imaging setting, it is required to install the red beam path. The near-infrared beam acts as the feedback control.}
    \label{top_view}
\end{figure}

For cross-polarization imaging, the directly reflected beam is automatically cancelled because a single reflection, ideally, does not rotate the polarization. 
Here, the incident beam needs to be linearly polarized. Here, we work with a p-polarized beam. The polarization is parallel to the optical table. 
This polarization is set using a polarizer in front of the laser. The illumination beam follows the beam path shown in Fig \ref{top_view}, however is guided through a polarizing beamsplitter instead of hitting the gold micro-mirror. P-polarization is reflected upwards, and s-polarization is transmitted. After laser-particle interaction, the back-reflected beam and the scattered beam share the same path and can not be distinguished. Arriving again at the polarizing beamsplitter, the beams are filtered by polarization; p-polarization is reflected and s-polarization is transmitted and collected by the camera. Because of the high-NA objective, the light is tightly focused on the sample and partially rotates the beam polarization. 
Direct reflection from a flat interface does not change the polarization, so it will be filtered out by the beamsplitter. 
This type of microscope is more sensitive to curved features on the interface, giving more contrast to the scattered light from the particle. 
If required, you can replace the high-NA objective with a lower NA objective and improve the field-of-view, at the cost of imaging resolution. 
Here, we make use of the total internal reflection for stabilization of the imaging plane, by installing a second light source. 
This second beam hits the glass-sample interface at an angle larger that total internal reflection, to separate it from the other coherent light source. 
It is aligned such that it hits the backside of the objective and is not going on the optical axis. 
Part of the out-going beam is directed through the polarizing beamsplitter again to be collected by the QPD. 
A dichroic mirror (Component No.44) separated the beams; the longer wavelength is transmitted and the shorter wavelength is reflected.

\begin{figure}[t]
    \centering
    \includegraphics[scale=0.25]{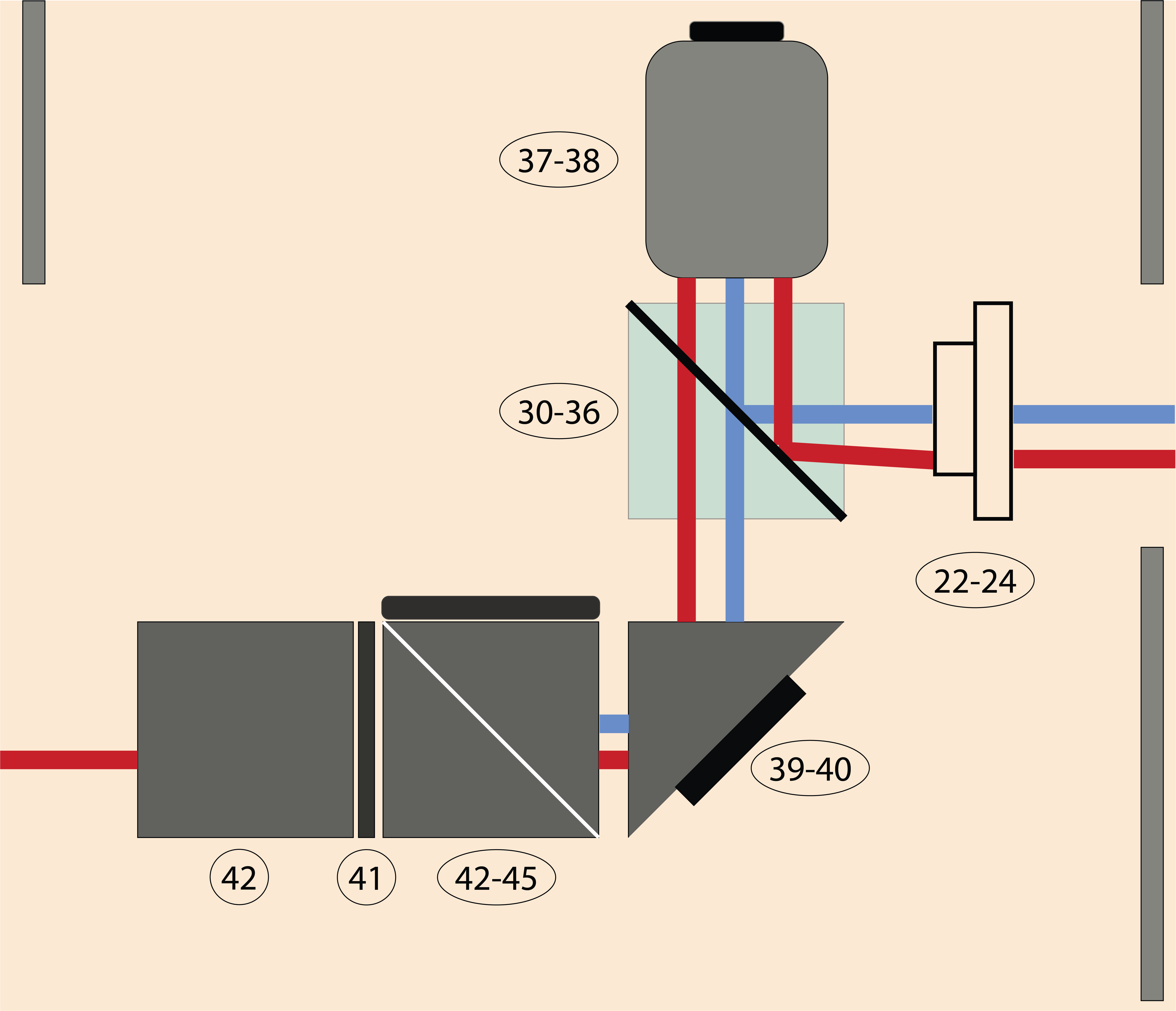}
    \caption{Enlarged image of orange colored region from Fig \ref{top_view} is shown when operating in cross-polarization. The beam-guiding region and parts of the feedback control are inside the aluminium skeleton (depicted as the grey walls on the outside of the image).}
    \label{side_view}
\end{figure}

\subsubsection*{Optical feedback control system}
Both setups have an integrated optical feedback control system, which increases the sensitivity and stability during measurements. The incoming laser beam hits the QPD sensor and detects the sum. The QPD difference signal corresponds to the actual position of the glass slide surface, and is fed to the to the NanoDrive as feedback on a PID signal. The NanoDrive has an in-built reader which translates voltage to $\mu m$ displacement. 
If you conduct your experiments in closed-loop configuration, the NanoDrive adjusts its position according to the signal it receives from the optical beam.

\subsubsection*{Image acquisition}
Microscope images are acquired by the Hamamatsu Orca Flash 4.0 sCMOS camera C11440-22CU. The camera runs at 30 fps at full size image if connected via USB 3.0 to the computer. 
Switching to Camera Link would increase the frame rate up to 100 fps at full size image 2048x2048 pixels.
\\\\
The current setup description guarantees reliable single-particle detection of submicrometer-sized particles in a 50x50 $\mu m$ field-of-view. Combining the hardware with the open-source software 'PyNTA' \cite{faez2019pynta}, single-particle tracking for long time periods is facilitated, and stores information about the particle's position (position coordinates), scattered intensity (pixel intensity of specified area), electronic output signal and camera trigger.

\section*{\textit{Design files}}

\subsubsection*{Micro mirror arm adapter}
This customized adapter is fabricated out of aluminium. Each metal arm is fixed on a compact kinematic mirror mount (Component No.36). On the front cutout, a triangular mirror is placed and tightened with a screw. Each mirror is coated with a high-reflective material e.g. gold. For extra secure hold, the mirrors are glued on their edges to the metal holder. Next, the base (Component No.35) needs to be prepared by cutting out an hole with a diameter of 20 $mm$. After assembling the arms, they are mounted at a certain distance apart on the base. The distance is necessary to bring the mirrors close or further apart from each other during the laser alignment. A gap in-between the mirrors allows the scattered beam to go through to the imaging path. This base has been chosen to fit the C6WR cage system from Thorlabs.

\begin{figure}[h!]
    \centering
    \subfloat[Front view.]{\includegraphics[width=0.3\textwidth]{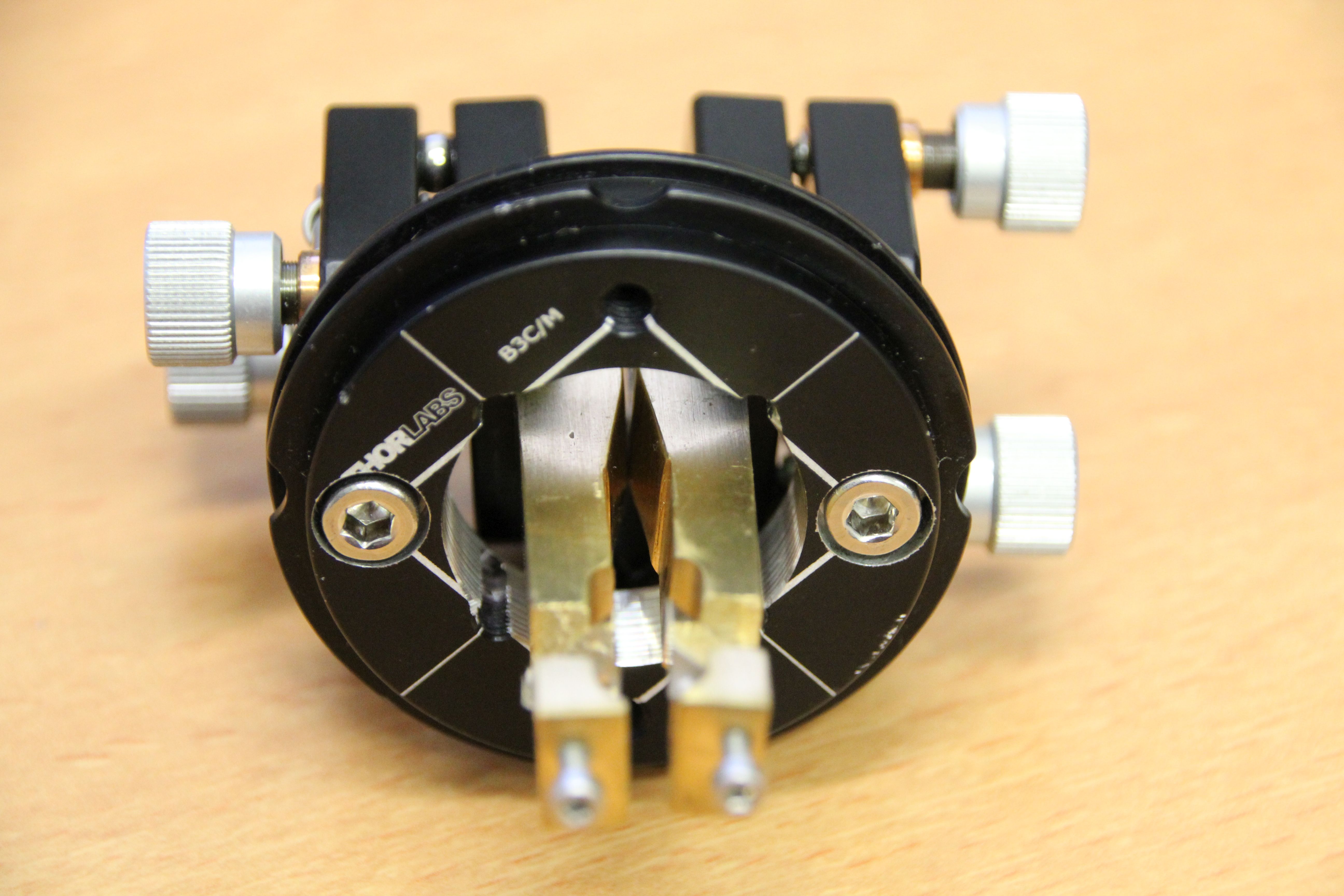}\label{fig_1}}\quad
    \subfloat[Top view.]{\includegraphics[width=0.3\textwidth]{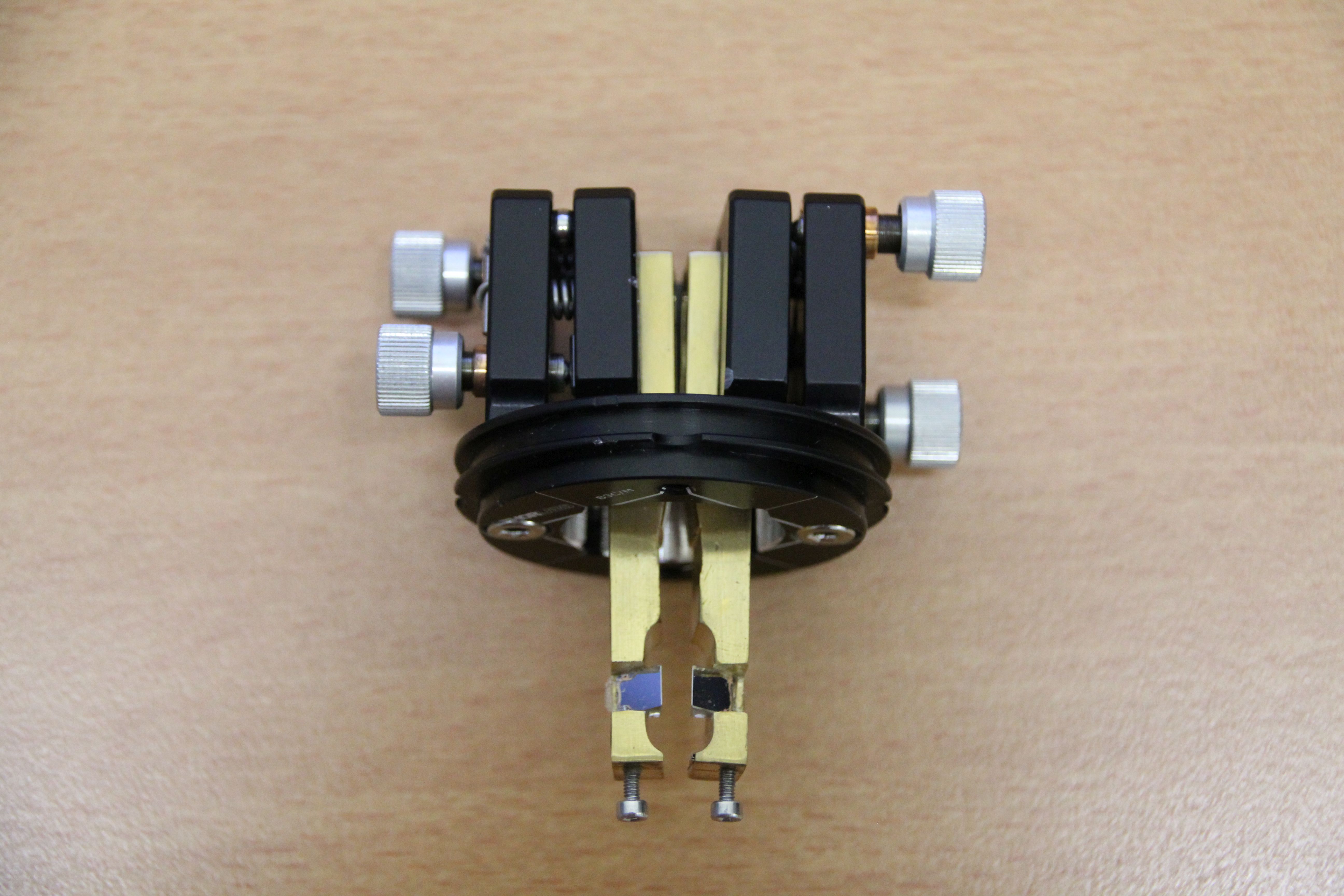}}\quad
    \subfloat[Side view.]{\includegraphics[width=0.3\textwidth]{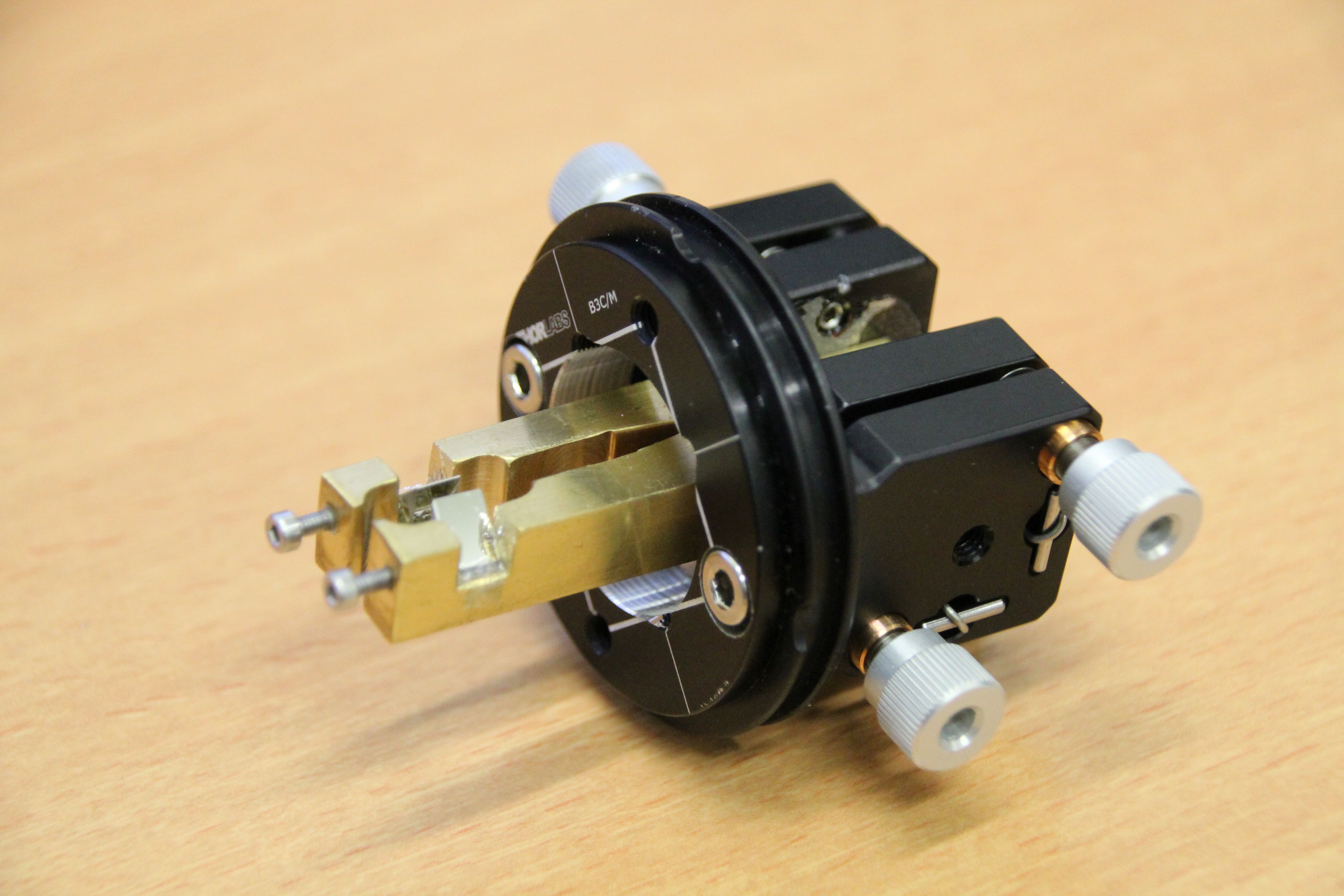}}\quad
\caption{Custom-built micro mirror arm adapter. The arms are made out of metal and are fixed on kinematic stages. At the front, the high-reflective mirror (mounted on the right arm) directs the incoming beam upwards. The reflected beam is guided by the second mirror (on the left arm) to the quadrant photodiode (QPD). That beam acts as the optical feedback control. The scattered beam propagates between the two arms. This design has been chosen to be compatible with the C6WR cage system from Thorlabs.}\label{micro_arm}
\end{figure}

\subsubsection*{Sample holder} 
The sample holder is made of aluminium to ensure hard contact onto the other metal parts. Glass slides with dimensions of 24x40 $mm$ fit exactly in the milled lowest region. 

\subsubsection*{Aluminium skeleton}
The aluminium skeleton protects the sensitive optical components from external influences such as dust particles, and air flow distortion, and serves as a base for the installation of the MicroDrive and NanoDrive stages. Precise cutouts on the walls are necessary that light can enter and leave the system.

\section{Design files summary}

\vskip 0.1cm
\tabulinesep=1ex
\begin{tabu} to \linewidth {|X|X|X[1.5,1]|X[1.5,1]|} 
\hline
\textbf{Design filename} & \textbf{File type} & \textbf{Open source license} & \textbf{Location of the file} \\\hline
SampleHolder \label{sample_holder} & CAD file & CC BY-SA 4.0 & \url{https://doi.org/10.17605/osf.io/t3yvd}  \\\hline
mirror\textunderscore arms & CAD file & CC BY-SA 4.0 & \url{https://doi.org/10.17605/osf.io/t3yvd} \\\hline
Aluminium\textunderscore skeleton & CAD file & CC BY-SA 4.0 & \url{https://doi.org/10.17605/osf.io/t3yvd} \\\hline
\end{tabu}

\section*{\textit{Bill of materials}}
A detailed list of all numbered components and associated costs can be found here \url{https://doi.org/10.17605/osf.io/t3yvd}.
\\\\
Please note that suppliers mentioned here are just suggestions. Material costs may vary over time and depend on location, and alternative sources may provide better prices than those listed here.
\\\\
We divided the setup into parts:
\begin{itemize}
\item[$\bullet$] Red beam path.
\item[$\bullet$] Blue beam path.
\item[$\bullet$] Option 1: Total internal reflection.
\item[$\bullet$] Option 2: Cross-polarization.
\item[$\bullet$] Feedback control.
\item[$\bullet$] Image acquisition.
\end{itemize}
as depicted in Fig \ref{top_view}. In the designated "Red beam path", the fiber-coupled laser beam can be replaced with a solid state laser improving the laser stability. Whether you choose for \underline{Option 1: Total internal reflection} \textbf{or} \underline{Option 2: Cross-polarization} microscope, most of the components are the same, but you need to switch the micro mirror arms with the polarising beamsplitter. The base (Component No.35) remains the same to fit the cage cube C6WR (Component No.30). In the designated "Image acquisition", you can choose your camera of choice. Here, we have installed a Hamamatsu Orca Flash 4.0.

\section{Build instructions} \label{build}
The framework builds up on Thorlabs Optical Cage System 30 mm metric units, more information can be found online under the link \url{https://www.thorlabs.com/navigation.cfm?guide_id=2002}.

\subsubsection*{Laser alignment }

\subsubsection*{Total internal reflection microscope}

Before you start aligning the illumination beam, you assure that every optical component is mounted and fixed as described in section \ref{hardware_description} and according to the steps explained in section \ref{build}. 
After leaving the fiber coupling output collimator (Component No.17), the beam hits the dichroic mirror (Component No.12) close to the center and is guided further to the next optical element No.14 by using the two knobs on the back of the first right-angle mirror mount (Component No.13, top right). 
The general rule of thumb is to adjust the blue light source to the center of each optical component. 
The second right-angle mirror mount has again two knobs on the backside to correct for the beam path. The light beam hits the center of the first mirror of the micromirror arm adapter (see Fig \ref{micro_arm}). 
Under optimal conditions, the beam should be reflected exactly under a 90$^\circ$ degree angle upwards. 
The light beam is guided on the backside of the objective and hits the glass slide under an oblique angle (more information provided in section \ref{hardware_description}). If you take a small piece of paper and hold it in a vertical position close to the hole on the sample holder, you should see a thin blue line. The reflected beam is leaving the system under the same incoming, oblique angle. 
Our next goal is to align the reflected beam such, that it will hit onto the QPD. You need to position the second mirror accordingly. 
On the backside of the micromirror arm adapter are two knobs per mirror, which you need to modify. Additionally, you need to guarantee that the space between the mirrors is large enough to pass a large fraction of the scattered light. 
\\ \break
\textit{Remark: You don't need the red beam path at all. The reflected blue laser beam can also act here as your optical feedback system.}

\subsubsection*{Cross-polarization microscope}
Before you start aligning the illumination beam, you assure that every optical component is mounted and fixed as described in section \ref{hardware_description}. After leaving the fiber coupling output collimator (Component No.17), the beam hits the dichroic mirror (Component No.12) close to the center and is guided further to the next optical element No.14 by using the two knobs on the back of the first right-angle mirror mount (Component No.13, top right). 
The general rule of thumb is to adjust the blue light source to the center of each optical component. 
The second right-angle mirror mount has again two knobs on the backside to correct for the beam path.
You remove the sample holder and unscrew the clean microscope objective (no residue of oil on the objective), and put them aside. 
Now, you have a clear view from top on the beam path. 
The light beam is reflected of the center of the polarizing beamsplitter (Component No.36). 
You can see the outcoming laser from top through the cutout of the sample holder. 
To check for the accuracy, you need to investigate the far-field laser beam-profile and its position. 
It is helpful to place a white sheet of paper at a large distance (roughly 1-2 meters) above your optical table and mark the center of the cutout. Again you use the two knobs on the second right-angle mirror mount (Component No.13, lower left) to guide the blue laser beam on the marked cross. 
Screwing the objective back at its position, should not alter the central position of the beam on your paper guideline (but will change its size). The p-polarization of the reflected beam should follow the same beam path as the incoming beam, because it is reflected on the polarizing beamsplitter. Only the s-polarized component of the scattered light is transmitted through the polarizing beamsplitter. 
This signal is relatively weak, preferably you can dim the lights or you already work in a dark room. 
After passing through the beamsplitter, it hits the elliptical mirror (Component No.40). 
An elliptical mirror is used to maintain symmetry (circular aperture) and reduce losses. 
The beam is further guided onto a shortpass dichroic mirror (Component No.44), which reflect the blue beam. Afterwards, the beam is going through two achromatic doublet lenses (Component No.59 \& 63) to minimize spherical aberration. 
Finally, the short wavelength beam reaches the camera. 
Wavelengths above 550 $\mu m$ are filtered out before entering the camera to cancel the near-infrared light used for stabilization.
For installing the optical feedback loop, you need a second coherent light source. In the schematic Fig \ref{top_view}, we used a fiber-coupled laser diode.
The red laser beam hits the center of dielectric mirror (Component No.4) and through the quarter-wave plate (Component No.6). We create a circularly polarized light, otherwise the beam can not pass through the polarizing beamsplitter in reflection. 
From now on, we align the red beam laser off-center of every optical element to spatially separate the illumination path from the optical feedback path. Another reason is the protection of the laser diode itself. As a reference, you can install an iris before the dichroic mirror (Component No.12) to know the exact position of the beam. 
The beam should hit the iris and later the dichroic mirror on the right of the center. 
Further it hits the dielectric mirror (Component No.14, lower left) again on the right of the center. 
Before reaching the polarizing beamsplitter, the beam is guided through the plano-convex lens (Component No.23) to focus the light further. Within the beamsplitter, the beam hits the lower end of the cube and is reflected upwards. 
The beam is guided on the backside of the objective, and leaves the sample under the same angle. 
Some part of the light is reflected back. 
\\ \break
\textit{Remark: Make sure that the reflected beam does not hit the onset of the fiber-coupled laser diode. It might damage the diode and deteriorate the laser power.} 
\\ \break
The other part is transmitted and guided through the elliptical mirror (Component No.40) and the dichroic mirror (Component No.44). The long-wavelength laser beam is transmitted and collected by the QPD (Component No.56).

\section{Operation instructions}
\begin{figure}[h!]
    \centering
    \includegraphics[scale=0.3]{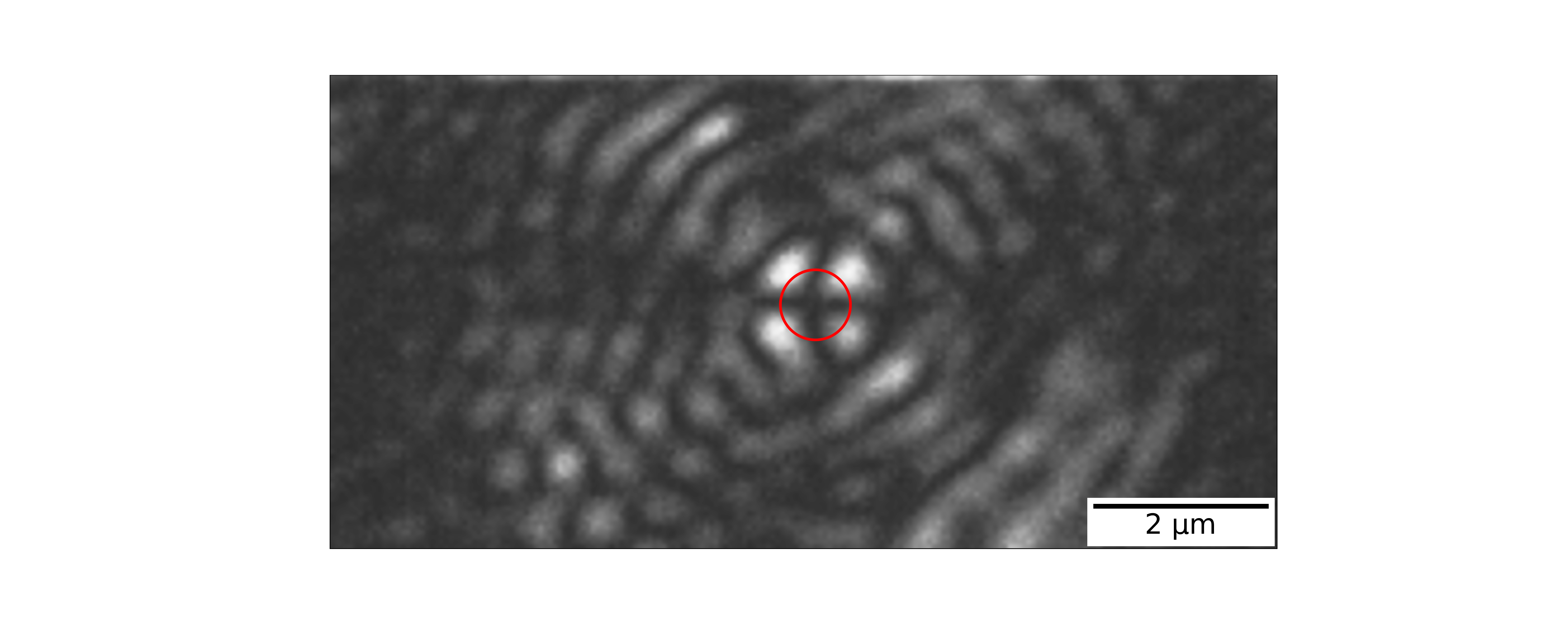}
    \caption{The point spread function (PSF) of a spherical polystyrene bead of 800 $nm$ in diameter, measured in cross-polarization mode. The red circle in the center marks the actual size of the particle and on the background, its diffraction pattern is shown. The characteristic clover-leaf pattern arises from this type of microscope.}
    \label{image_particle}
\end{figure}

Turn on the cooling system (air/water) of the camera and let it run for a while.
After a couple of minutes, you can start the camera by sliding the button. The light bulb next to the button should turn green. This indicates that everything works smoothly. If any other color code should appear, e.g. orange or red, read the manual or consult the company. Next, insert your flow cell or microfluidic chip into the sample holder. Be aware only microscope coverslips or other systems having the dimensions 24x40mm can fit into the sample holder. Fix the glass slide with magnets to ensure minimal drift. Before embedding the sample holder into its proper position, put sufficient (one to two drops) immersion oil on top of the objective. Keep in mind to use a proper amount of oil for index-matching. If you are using not enough oil, you will see the effect on the microscope image, resulting in a smaller field of view or no sight of any particle. Place the sample holder in its intended position on top of the objective. Based on the smearing of the oil on the glass, you can tell if you used the right amount or not. Turn on the blue laser, by enabling the button in its own software. If you need to adjust the power or other parameters, you can do so in the software itself. We keep the power low (1 mW) to lower the background scattering and not dealing with pixel saturation. To move around in the x- and y-plane, you need to operate the MicroDrive system (Component No.26). By adjusting the speed, you can regulate how fast to want to move within your system. After finding an object of interest, we use the adjustment knob next to the objective or the NanoDrive itself (Component No.27) to move along the z-direction and bringing the object into focus. As a reference, Fig \ref{image_particle} shows a spherical polystyrene bead of size: 800 $nm$ in focus while operating the cross-polarization microscope. 
The NanoDrive allows accurate displacement along the optical axis with sub-nanometer resolution.
We can turn on the red laser by tapping the ON button on its interface. You can change the power manually on the screen but keep in mind to stay in the low regime. The sum on the QPD should be still high enough, so that the received signal is readable. The object is now in focus and we attempt to bring as much reflected light from the red laser beam into the QPD by playing around with this two knobs on the backside of Component No.8. If we can ensure the position of the QPD signal at its center and a high SUM bar, we can fix the focal plane by switching from an open loop to a closed loop. It is important to find the correct settings for the PID loop by playing around first with the proportional gain, then the integral gain and at last the differential gain. We use an proportional gain of 0.42, an integral gain of 0.01 and a differential gain of 0.0, but it may vary for your experimental setup. In section \ref{validation}, you can find more information on how the stability in an open and closed loop configuration should look like. Try to minimize overcompensation by the PID loop, which will emerge as large oscillations on the display. You are ready to take some measurements. After aligning the whole setup, we recommend not to touch or modify any components in the illumination path because it will interfere with the rest of the setup and you might end up repeating all the steps above again.

\section{Validation and characterization}\label{validation}
\begin{figure}[h!]
\centering
\includegraphics[width=14cm]{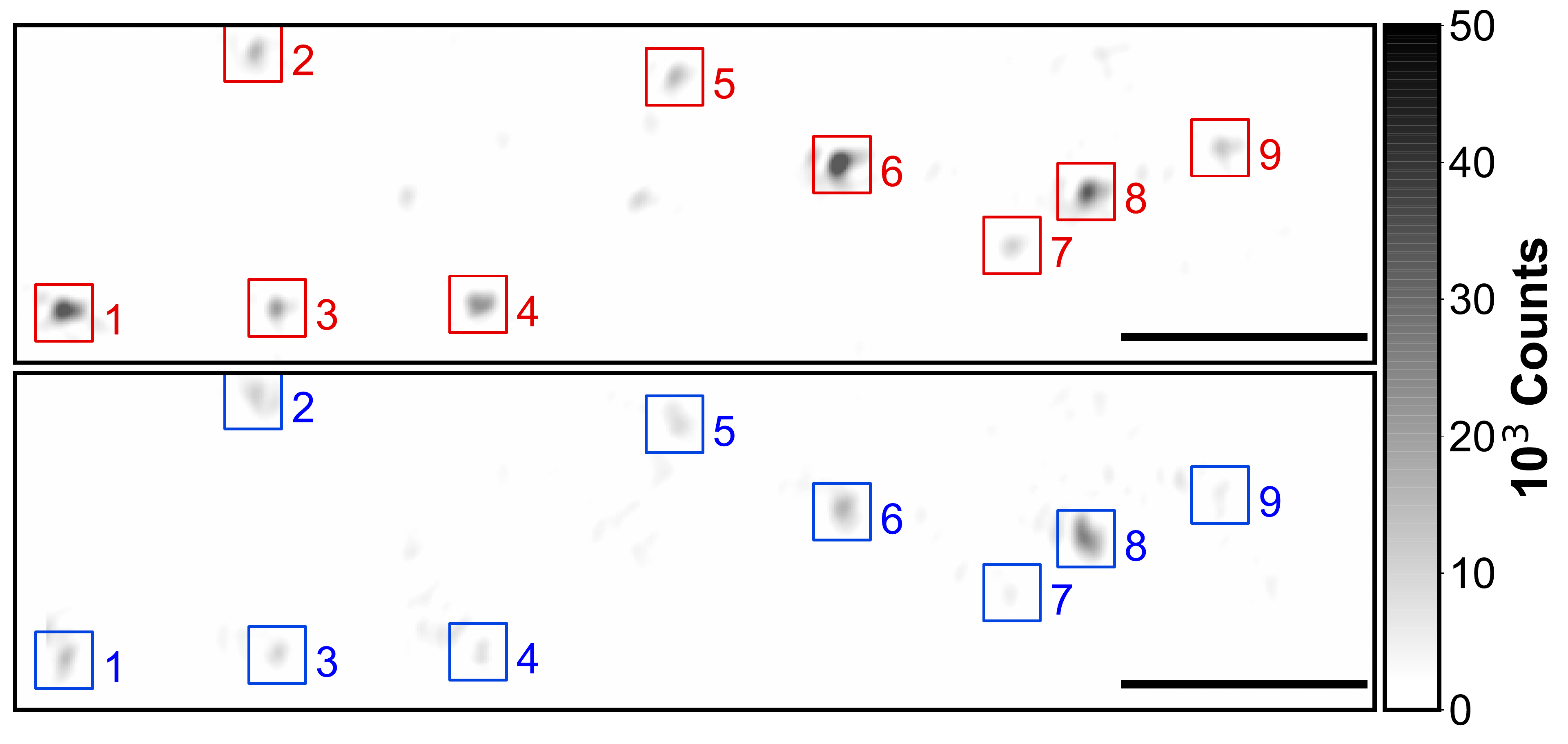}
\put(-390,80){b)}
\put(-390,170){a)}
\caption{The point spread function(PSF) of 30 $nm$ and 40 $nm$ nanoparticles(NPs) before and after stage drift in open loop. a) The particles landed on the glass slide before stage drift at time t=0 $min$. Before starting monitoring the stage drift over time, the NPs are brought into focus. b) The PSF of NPS after t=18 $min$ of drifting. The NPs are out of the focus plane of the objective. Scale bars: 4 $\mu$m}
\label{Fig_drift}
\end{figure}

\begin{figure}[h!]
\centering
\includegraphics[width=14cm]{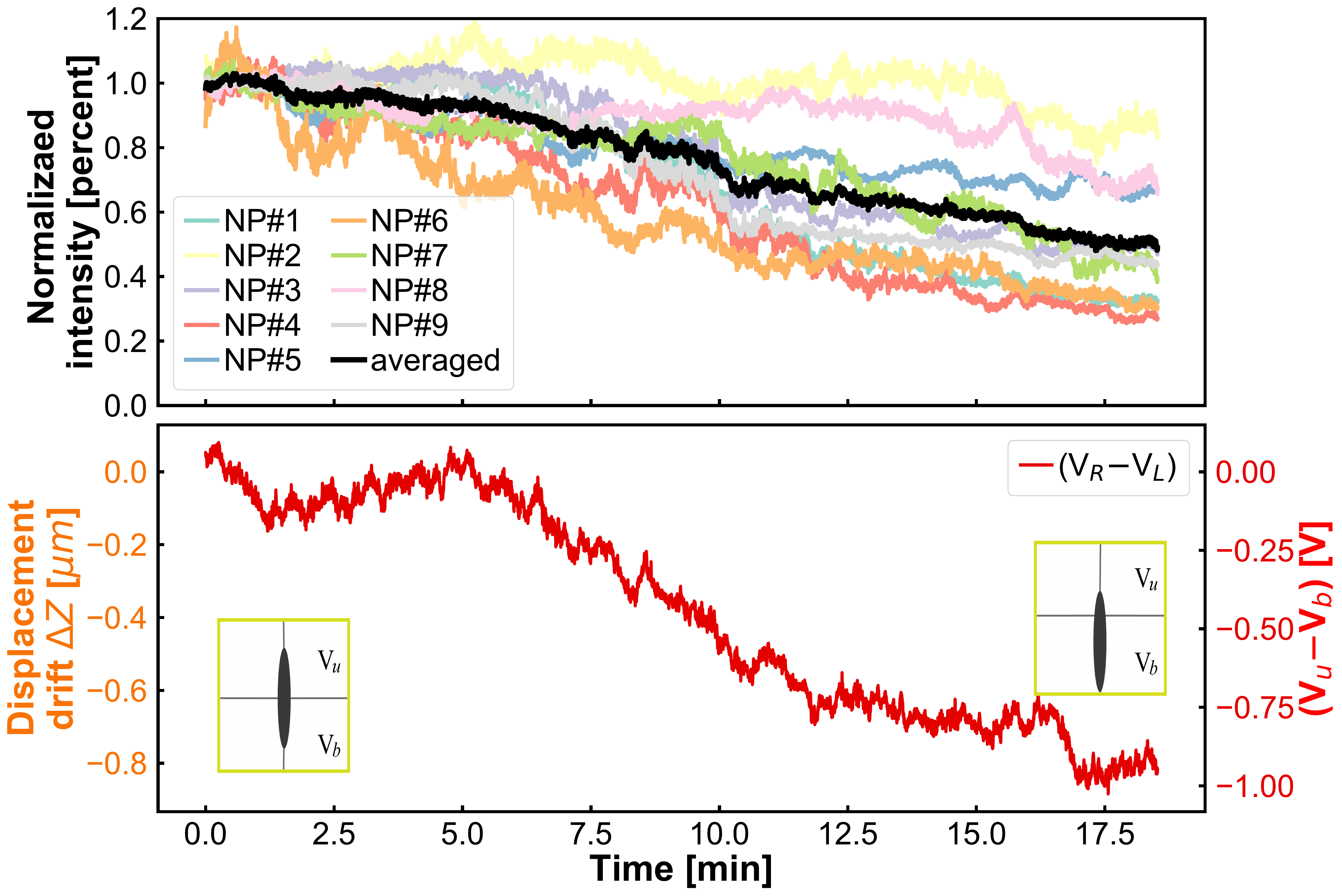}
\put(-440,130){b)}
\put(-440,250){a)}
\caption{The pixel intensity of each particle (out of nine from Fig \ref{Fig_drift}) has been recorded and normalized. All particles were measured simultaneously. The output voltage of the NanoDrive in open loop configuration is shown in the lower panel. After calibration, the voltage can be directly related to $\mu m$ displacement. a) The intensities of the NPs are not constant during the stage drifting. b) The differential voltage ($V_u - V_b$)  between the top and lower parts of the QPD during the stage drifting.}
\label{Fig_drift_intensity}
\end{figure}

\begin{figure}[h!]
\centering
\includegraphics[width=14cm]{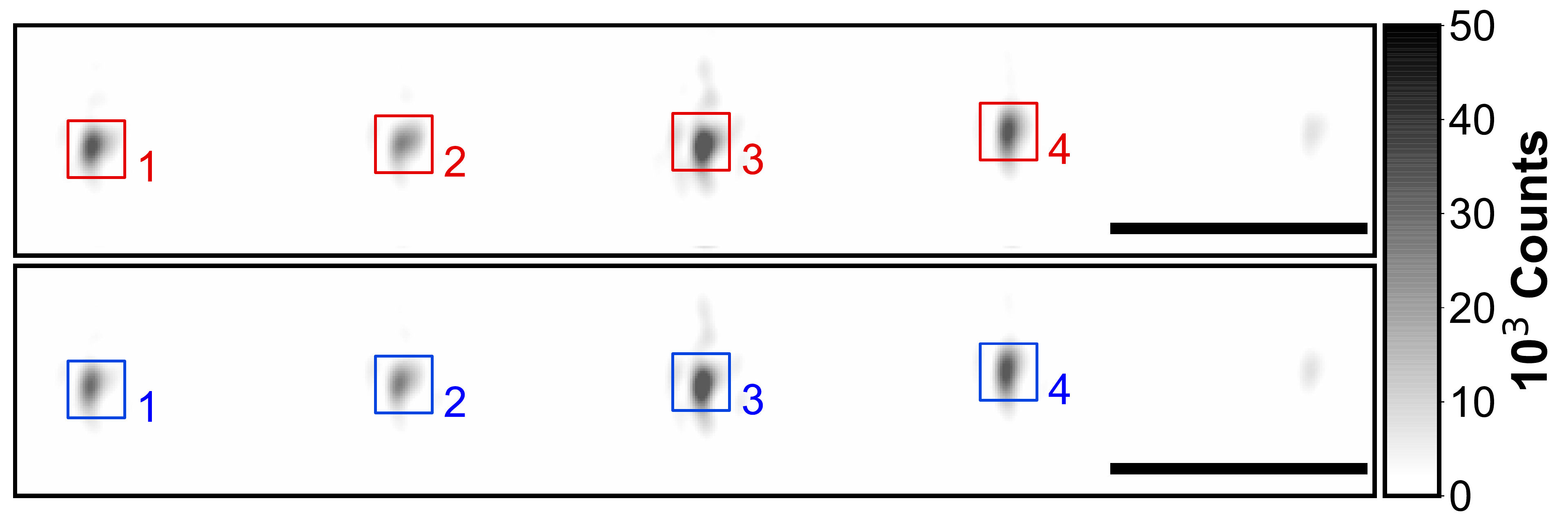}
\put(-390,55){b)}
\put(-390,115){a)}
\caption{PSF of NPs before and after 24-hour stage stabilization. The measurement regards 30 $nm$ and 40 $nm$ nanoparticles at the focal plane. a) The PSF of the NPs at the beginning of activating stage stabilization at time t=0 $min$. b) The PSF of NPS after 24 hours of activating stage stabilization. The NPs are still at the focus plane of the objective. Scale bar: 4 $\mu m$}
\label{Fig_stage_locki}
\end{figure}

\begin{figure}[h!]
\centering
\includegraphics[width=14cm]{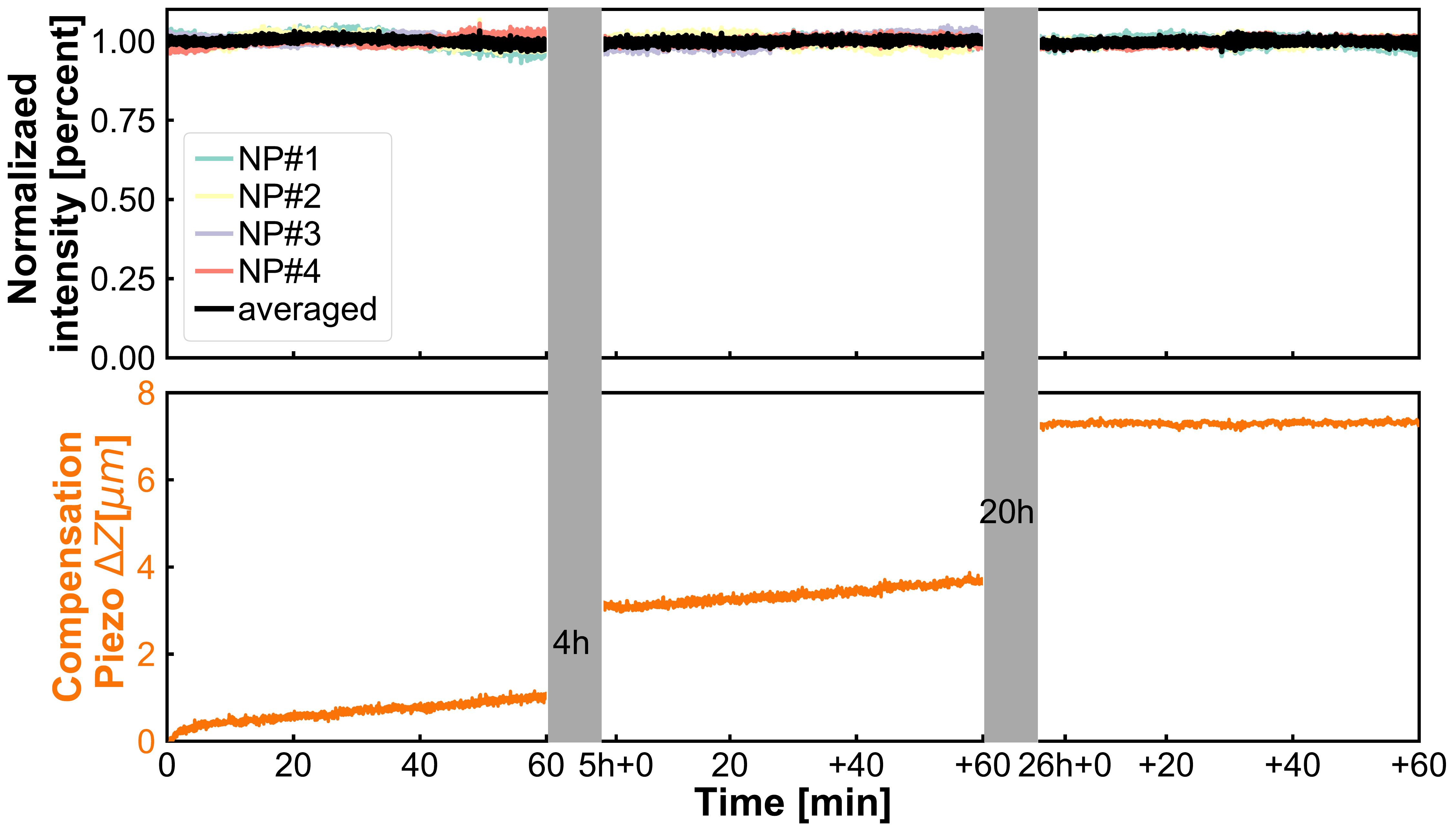}
\put(-440,110){b)}
\put(-440,210){a)}
\caption{a) NPs intensities tracks during activation of the stage Z-lock close loop control. The averaged intensity remains constant over time, indicating no observable stage drift. b) The displacement is compensated by the Piezo during the Z-lock close loop control. Left panel: first 60 minutes of monitoring right after activating the Z-lock. Middle panel: 60 minutes of monitoring 5 hours after activating the Z-lock. Right panel: 60 minutes of monitoring 24 hours after activating the Z-lock. }
\label{Fig_stage_lock}
\end{figure}

\begin{figure}[h!]
\centering
\includegraphics[width=14cm]{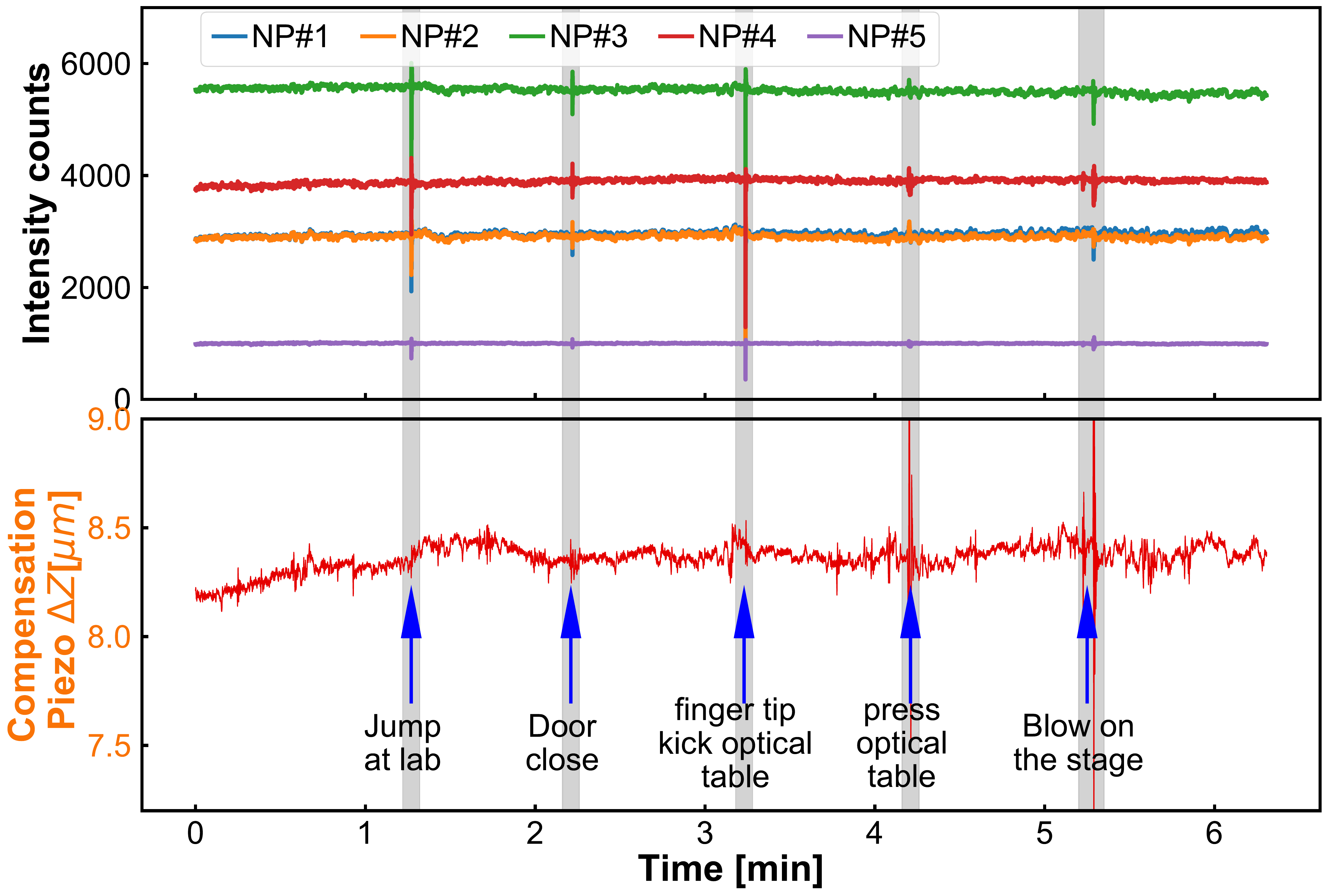}
\put(-440,130){b)}
\put(-440,260){a)}
\caption{\label{Fig_stage_vibration} a) Intensity tracks of NPs in a vibrations environment which is introduced artificial physical activities. b) The compensated displacement of piezo stage during this vibration environment.}
\end{figure}

We investigate the stability of the experimental set-up by analyzing how the point-spread function, the scattering intensity per nanoparticle and the QPD voltage change over time. All of the presented results below (Fig \ref{Fig_drift}-Fig \ref{Fig_stage_vibration}) are obtained with the total internal reflection microscopy setting. The feedback system allows two settings: open-loop and closed-loop. In open-loop, the optical signal is just monitored, but no input is send to the NanoDrive to correct for displacements along the optical axis of the objective. Hence in closed-loop, the microscope is continuously communicating with the NanoDrive via the QPD sensor, where the optical signal impinges. Figures \ref{Fig_drift} and \ref{Fig_drift_intensity} are obtained in open-loop. In a standard lab environment, a drift of roughly 0.8 $\mu m$ in 18 $min$ is presented. Upon switching to closed-loop, the drift can be reduced significantly (see Figures \ref{Fig_stage_locki} and \ref{Fig_stage_lock}).


\clearpage
\noindent
\textbf{Ethics statements}\\

\noindent
\textbf{CRediT author statement}
\\\\
{\textbf{Peters, Anna}: Visualization, Writing - Original Draft.  \textbf{Zhang Zhu}: Formal analysis, Validation. \textbf{Faez, Sanli}: Conceptualization, Supervision, Writing - Review \& Editing, Project administration, Funding acquisition. }
\\\\
\textbf{Acknowledgements}
\noindent
\\\\
We thank Paul Jurrius, JanBonne Aans, and Dante Killian for technical support.
\\\\
This project was supported by the Dutch Organisation for Scientific Research (NWO) with grant project: Photonics Translational Research – Medical Photonics (MEDPHOT).

\noindent
\bibliography{References}
\vskip 0.2cm
\noindent

\end{document}